\documentclass{article}

\usepackage[preprint]{neurips_2024}
\usepackage{dsfont}
\usepackage{makecell}

\usepackage[utf8]{inputenc} % allow utf-8 input
\usepackage[T1]{fontenc}    % use 8-bit T1 fonts
\usepackage{hyperref}       % hyperlinks
\usepackage{url}            % simple URL typesetting
\usepackage{booktabs}       % professional-quality tables
\usepackage{amsfonts}       % blackboard math symbols
\usepackage{nicefrac}       % compact symbols for 1/2, etc.
\usepackage{microtype}      % microtypography
\usepackage{xcolor}         % colors

\usepackage{enumitem}
\usepackage{natbib}
\usepackage{framed}
\usepackage{listings}
\usepackage{xcolor}
\usepackage{color}
\usepackage{multirow}
\usepackage{algorithm}
\usepackage{algorithmic}
\usepackage{amsmath}
\usepackage{amssymb}
\usepackage[most]{tcolorbox}
\usepackage{subfigure}
\usepackage{xspace}

\definecolor{codegreen}{rgb}{0,0.6,0}
\definecolor{codegray}{rgb}{0.5,0.5,0.5}
\definecolor{codepurple}{rgb}{0.58,0,0.82}
\definecolor{backcolour}{rgb}{0.95,0.95,0.92}

\lstdefinestyle{mystyle}{
    commentstyle=\color{brown},
    keywordstyle=\color{magenta},
    numberstyle=\tiny\color{codegray},
    stringstyle=\color{codepurple},
    basicstyle=\ttfamily\footnotesize,
    breakatwhitespace=false,         
    breaklines=true,                 
    captionpos=b,                    
    keepspaces=true,                 
    numbers=left,                    
    numbersep=5pt,                  
    showspaces=false,                
    showstringspaces=false,
    showtabs=false,                  
    tabsize=2
}

\newcommand{\tool}{\textsc{PerfCodeGen}\xspace}

\newcommand{\prompt}[1]{\begin{tcolorbox}[colback=white,%gray background
colframe=black,% black frame colour
arc=3mm, auto outer arc]#1\end{tcolorbox}}

\title{\tool: Improving Performance of LLM Generated Code with Execution Feedback}

\author{\hspace{-1.7mm}
{ 
Yun Peng\thanks{Work performed while interning at Salesforce Research}\hspace{0.5mm} \textsuperscript{1}\hspace{0.5mm},
Akhilesh Deepak Gotmare\thanks{Corresponding author: akhilesh.gotmare@salesforce.com} \hspace{0.5mm}\textsuperscript{2}\hspace{0.5mm},
Michael Lyu\textsuperscript{1}\hspace{0.5mm},
Caiming Xiong\textsuperscript{2}\hspace{0.5mm}} \\
\textbf{Silvio Savarese\textsuperscript{2}\hspace{0.5mm}, Doyen Sahoo\textsuperscript{2}\hspace{0.5mm}} \vspace{1.3mm} \\
\textsuperscript{1}The Chinese University of Hong Kong \\
\textsuperscript{2}Salesforce Research \\}

\begin{document}

\maketitle
\begin{abstract}
   Large Language Models (LLMs) are widely adopted for assisting in software development tasks, yet their performance evaluations have narrowly focused on the functional correctness of generated code. Human programmers, however, require LLM-generated code to be not only correct but also optimally efficient. We propose \tool, a training-free framework that enhances the performance of LLM-generated code by incorporating feedback based on runtime during test case execution into the self-refinement iterations. With \tool, we achieve speedups for a significantly higher proportion of problems compared to using the base LLM with sophisticated prompting techniques. Applied to open language models like Phi-3-mini, \tool achieves runtime efficiency comparable to prompting powerful closed models like GPT-4. We achieve state-of-the-art runtime efficiency on benchmarks such as HumanEval, MBPP, and APPS, frequently surpassing the ground truth reference solutions with \tool using GPT-3.5 and GPT-4. Additionally, we demonstrate the effectiveness of our approach in enhancing code quality across a range of open LLMs of varying sizes including Phi-3-mini, Llama 3 8B, Mixtral 8x7B, Command R, and Llama 3 70B.

\end{abstract}

\section{Introduction}\label{sec:intro}

Language Models (LMs) are now widely used for code completion (\cite{chen21evaluating, austin2021program, nijkamp2022codegen}), as well as for tasks like unit test case generation (\cite{https://doi.org/10.48550/arxiv.2207.10397}), bug fixing (\cite{yang2024sweagent}), feature addition (\cite{zhang2024autocoderover}), and other stages of software development (\cite{hong2023metagpt, qian2023communicative}). A recent Github survey (\cite{github_blog_2023}) underscores this rapid proliferation, estimating that 92\% of U.S. based developers are already using AI coding tools. However, despite this widespread adoption of Code LMs, evaluation has almost exclusively focused on functional correctness of generated code (\cite{hendrycksapps2021, li2022competition, puri2021codenet, liu2024your}), often overlooking key aspects of code quality. Runtime efficiency of a program, in particular, is a central consideration in software design decisions, due to its significant impact on user experience, serving costs and carbon footprint of a software application (\cite{landes1995treatment, chung2012non}). \cite{singhal2024nofuneval} provide a distinct evaluation of Code LMs by focusing on non-functional requirements such as efficiency, security and maintainability, revelaing that current models generally falter on these key quality metrics. This is particularly concerning as less experienced developers are more likely to accept AI-suggested code (\cite{dohmke2023sea}), often neglecting quality aspects, which subsequently places a significant burden on the code review process (\cite{gitclear2023copilot}).

Previous works (\cite{aman23learning} and \cite{garg2022deepperf}) have proposed fine-tuning LLMs to generate performance improvements for a given working program. However, this approach is challenging to scale due to the need for parallel training data in the code domain, which can be significantly more expensive to collect and manually validate than natural language data. Additionally, these methods require the availability of a functionally correct input program to optimize, which is not typically the case when writing code from scratch. Moreover, they do not leverage execution feedback from unit tests, which has been shown to be crucial in improving code correctness (\cite{le2022coderl}). This lack of execution feedback utilization is a notable limitation, as unit tests provide valuable insights into runtime behavior and performance characteristics.

% Prompting advances such as Chain of Thought (\cite{wei2022chain, nye2021show}) and Self-Refine (\cite{madaan2024self}) have been shown to enhance LLM output quality without requiring additional training. However, these approaches have limitations, including issues with hallucinations and sycophancies, and LLMs are known to be highly unreliable in refining their initial responses (\cite{huang2023large, laban2023you, sharma2023towards}). 

Prompting advances such as Chain of Thought (\cite{wei2022chain, nye2021show}) and Self-Refine (\cite{madaan2024self}) have enhanced LLM output quality without additional training, but limitations remain, including issues with hallucinations, sycophancies, and unreliability in refining initial responses (\cite{huang2023large, laban2023you, sharma2023towards}). To address these, several recent works (\cite{chen2023teaching, shinn2024reflexion, gou2023critic, welleck2022generating, sumers2023cognitive}) have proposed providing verbal feedback or grounding from an automatic tool or environment to the LLM during the refinement stage. However, these approaches often fail to assess or improve on quality aspects beyond mere functionality when generating code.% to improve its output quality, with an almost exclusive focus on code functionalitly.

\begin{figure}
    \centering    \includegraphics[width=1.0\textwidth]{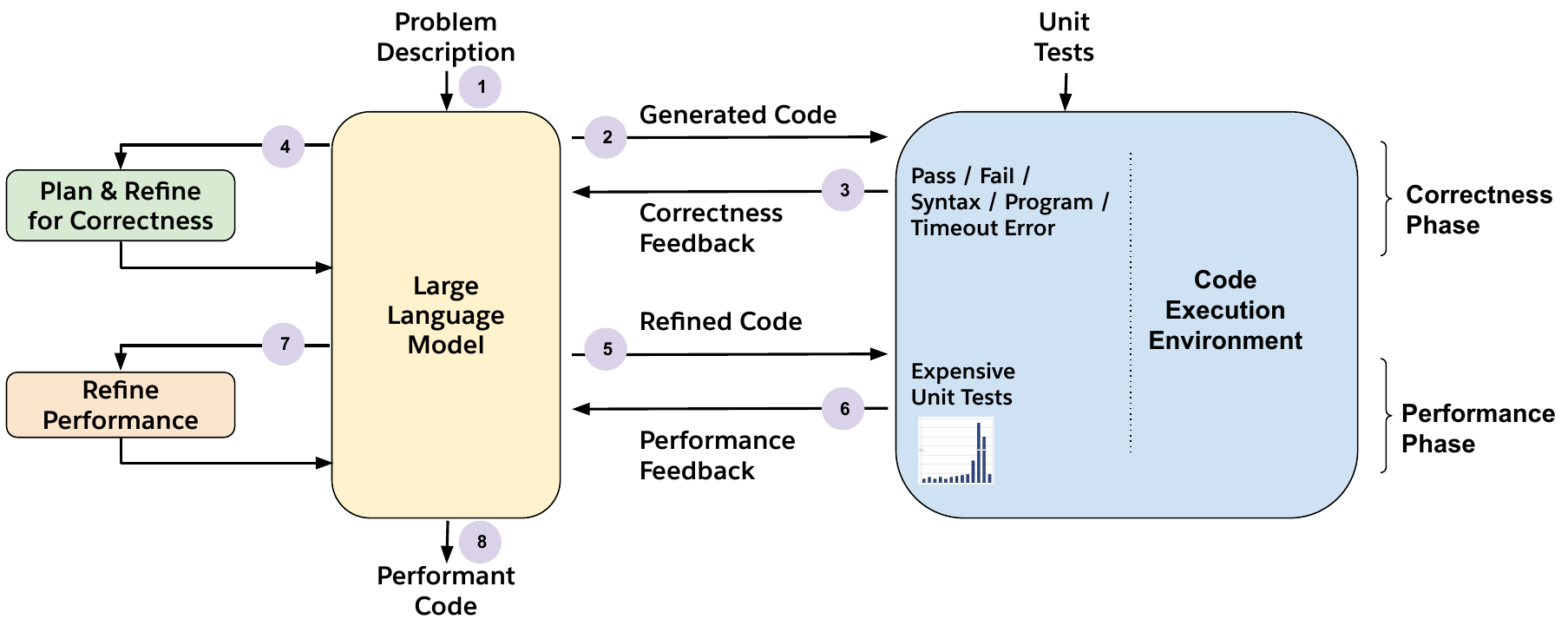}
    \caption{\tool Given a problem description ($1$), we prompt the LLM to generate a candidate solution ($2$), which is passed to an execution environment to collect feedback on correctness ($3$). The LLM is then prompted to self-reflect on this feedback in a planning stage, and accordingly generate a refinement using this context ($4$). This process is iterated over till correctness ($4$, $2$, $3$). Correct code obtained from this phase is self-refined for runtime-efficiency ($7$), and then passed to the environment to be executed ($5$) and the most time consuming unit test(s) is identified and passed as performance feedback to the LLM ($6$), that acts on it by generating a self-refinement to make the correct solution more efficient ($7$). This refinement is tested for correctness ($2$, $3$) and passed as the final code solution to the problem ($8$) if correct, else we fall back to correct program from ($3$) if any.}
    \label{fig:perfcodegen}
    % \vspace{-15mm}
\end{figure}

Inspired by this progress, we propose \tool, a framework that leverages code execution feedback in the iterative self-refinement stages of an LLM to improve the performance of generated code beyond merely ensuring functional correctness. First, we use environment feedback based on unit test execution to self-refine code generated from the base LLM for correctness. This yields a a larger set of correct solutions for our framework to optimize, increasing the likelihood of generating an optimally efficient program in the subsequent optimization phase. For performance refinement, we first assess the runtime required by correct solutions for each unit test, and then provide verbalised performance feedback indicating the most expensive unit test encountered during execution for the LLM to optimize its program. We evaluate \tool's effectiveness in generating runtime-efficient code on tasks from three widely used Python programming benchmarks: HumanEval (\cite{chen21evaluating}), MBPP (\cite{austin2021program}) and APPS \cite{hendrycksapps2021}. We use five open and two closed language models of varying sizes (Phi-3-mini 3.8B, Llama 3 8B, Mixtral 8x7B (13B), Command R 35B, Llama 3 70B, GPT-3.5 and GPT-4), and witness consistent and significant gains in the correctness and runtime efficiency of LLM-generated programs by applying \tool.

The remainder of this work is organized as follows: In Section \ref{sec:meth}, we provide a detailed explanation of the proposed \tool framework. Section \ref{sec:eval} presents experimental results demonstrating its effectiveness, accompanied by ablations comparing it with several other prompting approaches in Section \ref{subsec:ablation}. In Section \ref{sec:literature}, we discuss related work, followed by a brief conclusion in Section \ref{sec:conclusion}.
\section{\tool Methodology}\label{sec:meth}

We begin by prompting a given LLM with a problem description $x$ in natural language that details the requirements of the program to be generated. We sample $K$ candidate generations $C = \{y_x^i\}_{i = 1 \dots K} $ with nucleus sampling (using the base prompt from Figure \ref{fig:base}). As discussed in Section \ref{sec:intro}, we use the execution feedback to refine the incorrect programs within $C$ in order to increase the number of correct programs in this initial seed set for further optimization. We describe the correctness enhancements in Section \ref{subsec:correct_code}, and the performance refinement phase in Section \ref{subsec:optim_code}. The resulting \tool framework with its correctness and performance phases is illustrated in Figure \ref{fig:perfcodegen}. 

% , returns a set $C_{\text{correct}}$ where each element has been refined for performance.

% In the performance refinement stage, we use execution feedback from multiple runs on the unit tests, to improve runtime efficiency, described in Section \ref{subsec:optim_code}.

\subsection{Generating Correct Code}
\label{subsec:correct_code}
We assess the correctness of LLM generated programs using a set of $J$ unit tests $U(x) = \{ u^{j}_{x} \}_{j=1}^{J}$ corresponding to a problem $x$. After assessing correctness of all candidates, we iteratively refine incorrect ones based on execution feedback from the unit tests. For any $y_{x}^i \in C$ that fails any of the unit tests, we prompt the LLM again, this time incorporating the environment feedback for correctness verbalised as part of the prompt along with the failed solution and one of the failing unit test. The LLM is instructed to reflect and plan, and then generate a refined correct solution based on the reflection and planning result. The specific prompts we use are provided in Figure \ref{fig:correctness}.

% The LLM is more likely to generate a correct solution with the reflection-planning phase (analyzing the reason to fail and proposing a strategy to modify) included (as suggested by prior work \cite{wei2022chain, nye2021show}). 

The inclusion of a reflection-planning phase, as suggested by prior work \cite{wei2022chain, nye2021show}, increases the likelihood of the LLM generating a correct solution. The final refinement of $y_{x}^{i}$ generated by the LLM is then tested for correctness, and is passed for the performance optimisation stage if it passes all the unit tests. Otherwise, the problem $x$ is considered as unsolved for correctness by the given LLM. Subsequently, a set $C_{\text{correct}}$ is constructed by removing incorrect samples from $C$ and including their refined versions, if any of them achieve correctness. We could continue this iterative approach to further improve  the correctness rate of a given LLM, but to minimize computational costs, we pause after this one iteration. Besides, we observe that the gains in correctness significantly diminish after the first iteration of refinement. Note that this stage is shared across all the different performance refinement methods studied in this work. Having a larger number of correct candidates as seeds for performance refinements benefits the framework's effectiveness, as this implies higher likelihood of \tool generating an optimally efficient program.

\subsection{Optimising Correct Code using Execution Feedback}
\label{subsec:optim_code}

For all correct solutions $y_{x}^i \in C_{\text{correct}}$ that are constructed using the samples and refinements as described in Section \ref{subsec:correct_code}, we prompt the model to refine its solution to optimise performance while preserving functional equivalence. If this refinement breaks correctness, we stop and return the fastest program from $C_\text{correct}$. Otherwise, we measure the execution time consumed by this initial refinement (still denoted by $y^i_x$ for simplicity) to pass each available unit test $u_{x}^{j}$ corresponding to the given problem $x$. This involves conducting $E$ independent executions for each solution-test pair in identical compute environments. After sorting this set of $E$ observations, let $t(y^{i}_{x}, u^{j}_{x})[e]$ be the $e$-th smallest execution time consumed by $y^{i}_{x}$ on the $j$-th unit test $u^{j}_{x}$. We then calculate the empirical estimate of the expected execution time of a solution on a unit test, excluding the two outliers (smallest and largest execution times) to minimize the impact of potential measurement noise as follows: 
\begin{equation}
    \hat{t}(y^{i}_{x}, u^{j}_{x}) = \frac{1}{(E-2)} \sum_{e=2}^{E-1} t(y^{i}_{x}, u^{j}_{x})[e]; \hspace{6mm} u^{f}_{x} = \text{argmax}_{u^{j}_{x}} \hat{t}(y^{i}_{x}, u^{j}_{x}) 
    \label{equation:t_hat_estimate}
\end{equation}
We hypothesize that the $f$-th unit test $u^{f}_{x}$ of a problem $x$, that corresponds to the largest execution time, as defined above, can be highly informative in optimising the performance of $y_{x}^{i}$ if included in the feedback to the LLM for generating a revision. This approach mirrors how developers would identify the most time consuming pieces (hot spots) in their code to come up with runtime improving code changes. Therefore, we re-prompt the model with its latest generation ${y}_{x}^{i}$, its most time consuming unit test $u^{f}_{x}$ and an instruction to optimise the performance given this feedback (detailed in Figure \ref{fig:testcase_feedback}). This leads to a refinement denoted by $\tilde{y}_x^i$. The fastest amongst the refined correct outputs ($\{ \tilde{y}_x^i | \, \tilde{y}_x^i \text{  passes correctness} \}_{i=1 \dots K}$) is considered as the final performant solution for $x$. We employ the greedy decoding algorithm (sampling temperature set to $0$) in this stage of performance refinement, as we intend to collect only one refinement per correct code piece to minimize LLM inference costs. If none of the refinement $\tilde{y}_{x}^{i}$ pass correctness, we fall back to the fastest correct base solution from $C_{\text{correct}}$, where fastest from $C_{\text{correct}}$ is also found using $E$ executions on all unit tests. Similar to optimising for correctness, we could continue refining for performance with more iterations, but we pause after this one iteration for algorithmic simplicity and lower inference costs. Note that the LLM self-refinement step for runtime-efficiency is planned both before and after environment execution. To maximize the utility of execution feedback, we aim to expose to the environment a program that has undergone sufficient correctness and performance based self-refinement by the LLM. Ideally one could also include a planning substep in this self-refinement stage. However, to stay within the token limits of most LLMs, we omit this intermediate step. 
% , similar to the correctness phase
\section{Experiments}\label{sec:eval}

We describe the experiments demonstrating the effectiveness of \tool for generating runtime-efficient programs in this Section. Section \ref{subsec:eval_setup} outlines our experimental setup, Section \ref{subsec:eval_analysis} provides the main results of with all the models on the three benchmarks. In Section \ref{subsec:ablation} we compare \tool's execution feedback scheme with alternative prompting strategies, and in Section \ref{subsec:ablation_correct} we validate the effectiveness of the planning phase in improving correctness.

\subsection{Setup: Metrics, Datasets and Models}
\label{subsec:eval_setup}

To evaluate the correctness and runtime-efficiency of LLM generated solutions using different approaches, we follow prior work (\cite{aman23learning}) to compute and report the below metrics using the fastest (Best@$k$) LLM generated correct program out of $k$ samples.

\begin{itemize}[leftmargin=*]

    \item \textbf{Percent Optimized}  [\%Opt]: The proportion of problems in the test set where the fastest correct LLM generated program $y_x$ is more runtime-efficient (at least 10\% faster) than the ground truth $g_x$.
    
$ \frac{100}{N} \cdot \sum_{x}  \mathds{1}_{  \sum_{j}\hat{t} (y_x, u_x^j) < 0.9 \cdot \sum_{j}\hat{t} (g_x, u_x^j) } $  \text{where $N$ is the total number of test set problems}.
    
    \item \textbf{Percent Correct}  [\%Correct]: The proportion of problems in the test set where the LLM generates at least one correct solution out of $k$ candidates.

    \item \textbf{Speedup}: For problems where we obtain atleast one correct LLM-generated program $y_x$, we calculate speedup as the absolute ratio between the execution time required by $g_x$ and the execution time required by the fastest correct $y_x$. $\frac{ \sum_{x} \sum_{j}\hat{t} (g_x, u_x^j)}{\sum_{x} \sum_{j}\hat{t} (y_x, u_x^j)}$ %where $\sum_x$ only includes problems with at least one correct $y_x$

\end{itemize}

Following prior work (\cite{singhal2024nofuneval}), we rely on an empirical estimation of the execution time of Python programs, despite its drawbacks and challenges like high compute requirements. While tools like the gem5 simulator (\cite{binkert2011gem5}) for reliably and efficiently determine CPU cycles of a program, adapting them for Python is non-trivial. Nevertheless, our qualitative analysis (Listing \ref{lst:mbpp_main}) confirms that the differences observed in execution time ($\hat{t}$) correspond to clear differences in coding patterns. To estimate the execution time of a candidate solution, we use $E = 12$ executions for each unit test as described in Equation \ref{equation:t_hat_estimate}. We then compute the above three metrics using the fastest correct program $y_x$ obtained from $k$ (Best@$k$) candidates. If there are multiple ground truth solutions (like in APPS), we only use the fastest one as the reference for computing all metrics. We study the impact of sampling budget on the effectiveness of our framework by using $k \in \{1, 8, 20 \}$.
% set to either of $1$, $8$ or $20$ samples.
 % ($\sum_{j}\hat{t} (y_x, u_x^j)$)

We perform our analysis by treating \%Opt as the preferred metric over speedup when comparing different methods. A method achieving higher \%Opt would be generally more desirable to users than one with lower \%Opt, irrespective of speedups observed. This is because users often prefer a larger number of tasks solved optimally rather htan a few tasks solved more optimally. However, this preference can be adjusted based on user requirements in different contexts. Speedup should only be analyzed in conjunction with \%Opt and \%Correct, not in isolation, as it is defined using problems with a correctly generated program by the LLM. Note that the \%Correct metric is equivalent to the commonly reported pass@$k$, when $n = k$. However, since our approach leverages unit tests in execution feedback, it is not fair to compare our correctness metric with those from previous works (\cite{shinn2024reflexion}) that do not assume access to unit tests, and instead reserve them for correctness evaluation. Our study presents an alternative formulation where we seek to generate a program with optimal runtime efficiency, given all unit tests for a task. Our proposed framework can nonetheless be applied in settings we do not have any unit tests, by generating synthetic unit tests for a problem using LLMs (\cite{https://doi.org/10.48550/arxiv.2207.10397, shinn2024reflexion}). For simplicity, we instead re-purpose the HumanEval, MBPP and APPS datasets which include tests commonly used to evaluate LLMs on programming tasks. Appendix \ref{a:benchmark} details our pre-processing and dataset sanitisation steps. Given our focus on Python datasets, we are unable to compare with \cite{madaan2024self}'s CodeLlama 13B fine-tuned for generating C++ performance improvements. We include open LLMs of varying sizes: Phi-3-mini 3.8B (\cite{abdin2024phi}), Llama 3 8B (Meta), Mixtral 8x7B (13B active params, \cite{jiang2024mixtral}), Command R (Cohere), Llama 3 70B (Meta). We also include the closed and commercial GPT-3.5 (\cite{ouyang2022training}) and GPT-4 models via OpenAI APIs. We provide the details of our OpenAI API usage, LLM inference, and code execution environments in Appendix \ref{a:compute}.

\begin{table}
    \centering
    \scalebox{0.85}{
    \begin{tabular}{c c c c c c c c}
    % {c | c | c c c | c c c}
        \toprule
        \multirow{2}*{\textbf{Model}} & \multirow{2}*{\textbf{Method}} & \multicolumn{3}{c}{\textbf{Best@1}} & \multicolumn{3}{c}{\textbf{Best@8}} \\
        \cmidrule{3-8}
         & &  \textbf{\%Opt} & \textbf{\%Correct} & \textbf{Speedup} & \textbf{\%Opt} & \textbf{\%Correct} & \textbf{Speedup} \\
        \midrule
        \multicolumn{8}{c}{\textbf{HumanEval}} \\
        \midrule
        \multirow{2}*{Phi-3-mini} & Base & 14.63 & 51.83 & 1.24 & 35.98 & 78.05 & 1.44 \\
         & \tool & 18.29\tiny{\textcolor{blue}{(+3.66)}} & 57.32\tiny{\textcolor{blue}{(+5.49)}} & 1.23 & 40.85\tiny{\textcolor{blue}{(+4.87)}} & 85.37\tiny{\textcolor{blue}{(+7.32)}} & 1.68\\
         \midrule
         \multirow{2}*{Mixtral-8x7B} & Base & 9.43 & 27.04 & 1.31 & 19.5 & 63.52 & 1.37  \\
         & \tool & 11.32\tiny{\textcolor{blue}{(+1.89)}} & 32.70\tiny{\textcolor{blue}{(+5.66)}} & 1.31 & 27.67\tiny{\textcolor{blue}{(+8.17)}} & 75.47\tiny{\textcolor{blue}{(+11.95)}} & 1.71 \\
         \midrule
        \multirow{2}*{Command R} & Base & 19.02 & 54.6 & 1.37 & 25.15 & 71.17 & 1.43  \\
         & \tool & 20.25\tiny{\textcolor{blue}{(+1.23)}} & 57.06\tiny{\textcolor{blue}{(+2.46)}} & 1.37 & 32.52\tiny{\textcolor{blue}{(+7.37)}} & 79.75\tiny{\textcolor{blue}{(+8.58)}} & 1.46 \\
         \midrule
        \multirow{2}*{Llama 3 8B} & Base & 15.85 & 56.71 & 1.35 & 29.88 & 76.22 & 1.39 \\
         & \tool & 17.07\tiny{\textcolor{blue}{(+1.22)}} & 62.80\tiny{\textcolor{blue}{(+6.09)}}  & 1.29 & 31.10\tiny{\textcolor{blue}{(+1.22)}} & 81.71\tiny{\textcolor{blue}{(+5.49)}} & 1.62   \\ 
         \midrule
        \multirow{2}*{Llama 3 70B} & Base & 24.39 & 75.61 & 1.39 & 33.54 & 84.15 & 1.42  \\
         & \tool & 25.61\tiny{\textcolor{blue}{(+1.22)}} & 84.76\tiny{\textcolor{blue}{(+9.15)}} & 1.62 & 39.02\tiny{\textcolor{blue}{(+5.48)}} & 93.29\tiny{\textcolor{blue}{(+9.14)}} & 1.71 \\ 
         \midrule
        \multirow{2}*{GPT-3.5} & Base & 16.05 & 54.94  & 1.37 & 29.63 & 78.4 & 2.34  \\
         & \tool & 22.22\tiny{\textcolor{blue}{(+6.17)}} & 67.28\tiny{\textcolor{blue}{(+12.34)}} & 1.34 & 38.89\tiny{\textcolor{blue}{(+9.26)}} & 90.12\tiny{\textcolor{blue}{(+11.72)}} & 2.33  \\
         \midrule
        \multirow{2}*{GPT-4} & Base & 24.54 & 72.39 & 1.81  & 39.26 & 88.96 & 1.82 \\
         & \tool & 28.83\tiny{\textcolor{blue}{(+4.29)}} & 87.12\tiny{\textcolor{blue}{(+14.73)}} & 1.68 & 46.63\tiny{\textcolor{blue}{(+7.37)}} & 94.48\tiny{\textcolor{blue}{(+5.52)}}  & 1.91 \\
        \midrule
        \multicolumn{8}{c}{\textbf{MBPP (test)}} \\ %\hline
        \midrule
        \multirow{2}*{Phi-3-mini} & Base & 20.26 & 61.21 & 2.61 & 38.36 & 75.86 & 3.16 \\
         & \tool & 20.26\tiny{\textcolor{blue}{(+0.00)}} & 69.40\tiny{\textcolor{blue}{(+8.19)}} & 2.50 & 44.40\tiny{\textcolor{blue}{(+6.04)}} & 84.48\tiny{\textcolor{blue}{(+8.62)}}  & 3.03 \\
         \midrule
         \multirow{2}*{Mixtral-8x7B} & Base & 11.26 & 45.95  & 1.46 & 22.07 & 66.67 & 1.63 \\
         & \tool & 13.06\tiny{\textcolor{blue}{(+1.8)}} & 53.15\tiny{\textcolor{blue}{(+7.2)}} & 1.34 & 38.29\tiny{\textcolor{blue}{(+16.22)}} & 78.38\tiny{\textcolor{blue}{(+11.71)}} & 2.83  \\
         \midrule
        \multirow{2}*{Command R} & Base & 15.86 & 55.51 & 1.68 & 25.11 & 69.16 & 2.27 \\
         & \tool & 17.18\tiny{\textcolor{blue}{(+1.32)}} & 56.83\tiny{\textcolor{blue}{(+1.32)}} & 1.73 & 31.72\tiny{\textcolor{blue}{(+6.61)}} & 75.33\tiny{\textcolor{blue}{(+6.17)}} & 2.78  \\
         \midrule
        \multirow{2}*{Llama 3 8B} & Base & 16.02 & 59.74 & 2.21 & 28.14 & 71.86 & 2.23\\
         & \tool & 17.75\tiny{\textcolor{blue}{(+1.73)}} & 68.40\tiny{\textcolor{blue}{(+8.66)}} & 1.91 & 32.90\tiny{\textcolor{blue}{(+4.76)}} & 82.68\tiny{\textcolor{blue}{(+10.82)}} & 3.03  \\ 
         \midrule
        \multirow{2}*{Llama 3 70B} & Base & 18.92 & 65.32  & 1.44 & 26.13 & 77.93 & 1.63 \\
         & \tool & 17.12\tiny{\textcolor{red}{(-1.8)}} & 68.47\tiny{\textcolor{blue}{(+3.15)}} & 1.68 & 34.68\tiny{\textcolor{blue}{(+8.55)}} & 88.29\tiny{\textcolor{blue}{(+10.36)}} & 2.21 \\ 
         \midrule
        \multirow{2}*{GPT-3.5} & Base & 44.1 & 66.38 & 3.42 & 57.21 & 76.86 & 3.69  \\
         & \tool & 47.16\tiny{\textcolor{blue}{(+3.06)}} & 73.36\tiny{\textcolor{blue}{(+6.98)}} & 4.19 & 71.18\tiny{\textcolor{blue}{(+13.97)}} & 88.21\tiny{\textcolor{blue}{(+11.35)}} & 4.13  \\
         \midrule
        \multirow{2}*{GPT-4} & Base & 20.87 & 72.17 & 2.76 & 43.48 & 82.17 & 2.85\\
         & \tool & 30.87\tiny{\textcolor{blue}{(+10)}} & 86.52\tiny{\textcolor{blue}{(+14.35)}} & 2.70 & 56.52\tiny{\textcolor{blue}{(+13.14)}} & 93.91\tiny{\textcolor{blue}{(+11.74)}} & 4.01   \\ %\hline
        \bottomrule \\
    \end{tabular}}
    \caption{\%Correct, \%Opt, and Speedup results on HumanEval and MBPP with $k$ of $1$ and $8$. Gains in \%Opt and \%Correct observed by applying our \tool framework are indicated in the relevant cells (\textcolor{blue}{$+\delta$} , \textcolor{red}{$-\delta$} ) for each model. \tool generates correct and runtime-efficient programs for more problems than the base LLM. Despite the higher correctness rate, \tool maintains similar speedup, i.e. speedup over more samples, demonstrating its ability to generate runtime-efficient solutions for more challenging problems that the base LLM could not solve optimally.
    }
    \label{tab:main_result_he_mbpp}
    \vspace{-5mm}
\end{table}

\subsection{\tool Results}
\label{subsec:eval_analysis}

We report the performance of different LLMs on problems from the the HumanEval and MBPP benchmarks in Table \ref{tab:main_result_he_mbpp} using our \tool framework and the aforementioned metrics, with a sampling budget of $k = 1$ and $8$ samples. For comparsion, we also list the base LLM performance without using \tool. Performance results with a $k$ of $20$ are provided in Table \ref{tab:main_result_best_20} in the Appendix. On both the benchmarks, we witness that our framework leads to significant improvements in \%Opt and \%Correct for all base LLMs, particularly with the higher $k$ of $8$.

With \tool we notably enhance the runtime-efficiency of programs generated by open models like Phi-3-mini (\%Opt of $40.85$ @ $k = 8$) and Llama 3 70B  ($39.02$) making them comparable to GPT-4 in the base setting, which attains the highest base \%Opt of $39.26$ ($k = 8$). Similarly, with \tool, open models like Mixtral ($27.67$), Command R ($32.52$) and Llama 3 8B ($31.10$) can achieve comparable \%Opt to GPT-3.5 ($29.63$). While we can elevate the performance of open models to match that of closed commercial models, we witness even higher gains in \%Opt and \%Correct when using \tool on the closed GPT-3.5 and GPT-4 models. This finding can be attributed to the differences in the reasoning capabilities of these model categories, as the effectiveness of \tool heavily relies on the reasoning skills of the given LLM.

On MBPP, we continue to witness significant gains in \%Opt when using \tool in most cases when sampling $8$ candidates. An exception is Llama 3 70B, whose performance marginally drops on MBPP with our method at $k = 1$, likely due to the high variance in estimating \%Opt with a single sample. This drop can be mitigated in practice by leveraging execution time evaluation to fall back to the base LLM output if our refinement is correct but suboptimal. However, we avoid doing so here for a stricter evaluation of our scheme. We provide an example in Listing \ref{lst:mbpp_main} where \tool generates a faster program than the ground truth. As shown in Table \ref{tab:main_result_he_mbpp}, most LLMs generate solutions that are faster than the ground truth for significant portions of the test set, raising questions about their optimality.

\begin{lstlisting}[language=python,caption=An optimal solution generated by \tool in MBPP.,label=lst:mbpp_main]
'''
Problem: Write a python function to find the average of cubes of first n natural numbers.
'''
# Optimal solution generated by \tool based on GPT-4:
def solution(n):
    sum_of_cubes = (n * (n + 1) / 2.0) ** 2
    return sum_of_cubes / n
    
# Original ground truth solution in MBPP:
def solution(n):
    sum = 0
    for i in range(1, n + 1):
        sum += i * i * i
    return round(sum / n, 6)
\end{lstlisting}

\begin{table}
    \centering
    \scalebox{0.85}{
    \begin{tabular}{c c c c c c c c}
        \toprule
        \multirow{2}*{\textbf{Model}} & \multirow{2}*{\textbf{Method}} & \multicolumn{3}{c}{\textbf{Best@1}} & \multicolumn{3}{c}{\textbf{Best@8}} \\
        \cmidrule{3-8}
         & & \textbf{\%Opt} & \textbf{\%Correct} &  \textbf{Speedup} & \textbf{\%Opt\%} & \textbf{\%Correct} & \textbf{Speedup}  \\
        \midrule
        \multicolumn{8}{c}{\textbf{APPS (test)}} \\ %\hline
        \midrule
        \multirow{2}*{Phi-3-mini} & Base & 0.09 & 6.51 & 1.03 & 0.28 & 13.77 & 1.86  \\
         & \tool & 0.09\tiny{\textcolor{blue}{(+0.0)}} & 8.40\tiny{\textcolor{blue}{(+1.89)}} & 1.00 & 0.40\tiny{\textcolor{blue}{(+0.12)}} & 16.73\tiny{\textcolor{blue}{(+2.96)}} & 2.02  \\
         \midrule
        \multirow{2}*{Mixtral-8x7B} & Base & 0.12 & 6.33 & 0.97 & 0.31 & 11.02 & 1.08  \\
         & \tool & 0.19\tiny{\textcolor{blue}{(+0.07)}} & 7.08\tiny{\textcolor{blue}{(+0.75)}} & 1.95 & 0.40\tiny{\textcolor{blue}{(+0.09)}} & 14.50\tiny{\textcolor{blue}{(+3.48)}} & 1.81   \\
         \midrule
        \multirow{2}*{Command R} & Base & 0.31 & 14.27  & 1.00 & 0.58 & 24.86 & 1.00  \\
         & \tool & 0.43\tiny{\textcolor{blue}{(+0.12)}} & 16.48\tiny{\textcolor{blue}{(+2.21)}} & 1.00 & 0.86\tiny{\textcolor{blue}{(+0.28)}} & 28.82\tiny{\textcolor{blue}{(+3.96)}} & 1.00   \\
         \midrule
        \multirow{2}*{Llama 3 8B} & Base & 0.18 & 9.38 & 2.29 & 0.52 & 15.8 & 3.07 \\
         & \tool & 0.25\tiny{\textcolor{blue}{(+0.07)}} & 10.21\tiny{\textcolor{blue}{(+0.83)}}  & 2.62 & 0.61\tiny{\textcolor{blue}{(+0.09)}} & 17.71\tiny{\textcolor{blue}{(+1.91)}}  & 3.07  \\ 
         \midrule
        \multirow{2}*{Llama 3 70B} & Base & 0.37 & 18.65  & 1.04 & 0.65 & 26.12 & 1.04 \\
         & \tool & 0.31\tiny{\textcolor{red}{(-0.06)}} & 19.30\tiny{\textcolor{blue}{(+0.65)}}  & 1.04 & 0.86\tiny{\textcolor{blue}{(+0.21)}} & 27.94\tiny{\textcolor{blue}{(+1.82)}} & 1.04   \\ 
         \midrule
        \multirow{2}*{GPT-3.5} & Base & 0.49 & 25.18 & 1.33  & 1.11 & 39.92 & 1.25\\
         & \tool & 0.58\tiny{\textcolor{blue}{(+0.09)}} & 30.56\tiny{\textcolor{blue}{(+5.38)}} & 1.32 & 1.48\tiny{\textcolor{blue}{(+0.37)}} & 46.26\tiny{\textcolor{blue}{(+6.34)}} & 1.24   \\
         \midrule
        \multirow{2}*{GPT-4} & Base & 0.31 & 36.63  & 0.98& 1.14 & 57.75 & 1.15  \\
         & \tool & 0.61\tiny{\textcolor{blue}{(+0.30)}} & 45.62\tiny{\textcolor{blue}{(+8.99)}} & 1.26 & 1.96\tiny{\textcolor{blue}{(+0.82)}} & 65.80\tiny{\textcolor{blue}{(+8.05)}} & 1.90  \\
        \bottomrule \\
    \end{tabular}}
    \caption{\%Correct, \%Opt, and Speedup results on APPS with $k$ of $1$ and $8$. Gains in \%Opt and \%Correct observed by applying our \tool framework are indicated in the relevant cells (\textcolor{blue}{$+\delta$} , \textcolor{red}{$-\delta$} ) for each model. \tool generates correct and runtime-efficient programs for more problems than the base LLM. Despite the higher relative correctness rate, \tool maintains similar speedup, i.e. speedup over more samples. While the absolute \%Opt and \%Correct on APPS problems are lower due to the higher difficulty level of problems, \tool offers sizeable relative gains on both these metrics.
    }
    \label{tab:main_result_apps}
    \vspace{-5mm}
\end{table}

Results on the APPS dataset are reported in Table \ref{tab:main_result_apps}.
We observe a significantly lower correctness rate compared to HumanEval and MBPP with all LLMs, which is expected given the higher difficulty of APPS. Additonally, since each problem in this dataset has on average more than $25$ ground truth solutions, and we use the best one as the reference for comparison, most models fail to generate a solution more optimal than the best ground truth reference for over $1\%$ of the dataset with a $k$ of $1$. Using \tool's execution feedback, we observe gains in \%Opt, \%Correct and speedup of GPT-3.5 and GPT-4 generations on APPS-test problems, whereas open models benefit less due to task difficulty and their limited reasoning skills. Notably, the gap in correctness rates between commercial GPT models and open models is significantly wider on APPS than on simpler problems from HumanEval and MBPP, highlighting the capability differences.

% in their reasoning and programming capabilities.

\begin{table}[t]
    \centering
    \scalebox{0.9}{
    \begin{tabular}{c c c c c}
        \toprule
        \multirow{2}{*}{\makecell{\textbf{Prompting Method} \\ Summarized Instruction(s)}} & \multicolumn{2}{c}{\textbf{HumanEval}} & \multicolumn{2}{c}{\textbf{MBPP (test)}} \\
        \cmidrule{2-5}
         & \textbf{Speedup} & \textbf{\%Opt} & \textbf{Speedup} & \textbf{\%Opt} \\
        \midrule
        \makecell{\textbf{Base LLM Generation} \\ GPT-3.5 prompted to solve a problem} & 1.37 & 16.05 & 3.42 & 44.1 \\ \\
        \makecell{\textbf{Perf Improvement Prompt} \\ Optimize given code, maintaining equivalence} & 1.39 & 19.14\tiny{\textcolor{blue}{(+3.09)}} & 3.45 & 44.1\tiny{\textcolor{blue}{(+0.00)}} \\ \\
        \makecell{\textbf{In-context learning} \\ + Here's an example of optimisation:  \textit{\{demo\}}} & 1.38 & 19.14\tiny{\textcolor{blue}{(+3.09)}} & 3.20 & 44.98\tiny{\textcolor{blue}{(+0.88)}} \\ \\
        \makecell{\textbf{Pre-defined Strategies} \\ + Here are common ways to optimize: \textit{\{strategies\}}} & 1.27 & 16.67\tiny{\textcolor{blue}{(+0.62)}} & 2.86 & 32.75\tiny{\textcolor{red}{(-11.35)}} \\ \\
        \makecell{\textbf{Plan and Refine} \\ (a) Generate optim. plan (b) Optimize w.r.t. plan} & 1.34 & 19.14\tiny{\textcolor{blue}{(+3.09)}} & 3.30 & 45.85\tiny{\textcolor{blue}{(+1.75)}} \\ \\
        \makecell{\textbf{Analyze and Refine} \\ (a) Analyze $\mathcal{O}$ time (b) Optimize given a's analysis} & 1.35 & 18.52\tiny{\textcolor{blue}{(+2.47)}} & 3.32 & 46.29\tiny{\textcolor{blue}{(+2.19)}} \\ \\
        \makecell{\textbf{Multi-Agent w/ Reviewer} \\ (a) Coder: Optimize (b) Reviewer: Suggest changes} & 1.25 & 18.52\tiny{\textcolor{blue}{(+2.47)}} & 3.57 & 41.48\tiny{\textcolor{red}{(-2.62)}} \\ \\
        \makecell{\textbf{Multi-Agent w/ Team} \\ (a) Leader: Plan optim (b) Coder Optimize \\w.r.t. plan (c) Reviewer: Suggest changes to b \\} & 1.28 & 20.37\tiny{\textcolor{blue}{(+4.32)}} & 3.20 & 42.79\tiny{\textcolor{red}{(-1.31)}} \\ \\
        \makecell{\textbf{Direct Execution Feedback} 
        \\ (a) Optimize (b) It worked/didn't, try again
        % \& test performance 
        % \\ (b) Relay feedback to refine
        } & 1.38 & 21.60\tiny{\textcolor{blue}{(+5.55)}} & 3.58 & 42.79\tiny{\textcolor{red}{(-1.31)}} \\ \\
        \makecell{\textbf{\tool} \\
        (a) Optimize given code (b) Your costliest \\ unit test is \textit{\{ test \}}, optimize accordingly} & 1.34 & \textbf{22.22}\tiny{\textcolor{blue}{(+6.17)}} & 4.19 & \textbf{47.16}\tiny{\textcolor{blue}{(+3.06)}} \\
        \bottomrule \\
    \end{tabular}}
    \caption{Comparison of \tool Alternatives in the Performance-Refinement Stage ($k = 1$) using GPT-3.5. Gains in \%Opt are denoted by \textcolor{blue}{$+\delta$} or \textcolor{red}{$-\delta$} in the respective cells. \tool's execution feedback demonstrates the highest gains while maintaining a similar speedup.}
    \label{tab:ablation_prompting}
    \vspace{-3mm}
\end{table}

\subsection{Evaluating Alternatives to \tool's Execution Feedback}
\label{subsec:ablation}

Given a correct solution, we evaluate some common prompting techniques for the performance improvement phase. We provide the specific prompts in verbatim in Appendix \ref{a:prompts}. Table \ref{tab:ablation_prompting} lists the results with all these methods on the HumanEval and MBPP datasets using GPT-3.5. We exclude other models from this analysis to avoid the high costs associated with LLM inference. Due to the significantly larger size of the APPS-test set compared to HumanEval and MBPP, we exclude it from this comparative analysis to limit inference costs. APPS-test contains roughly $20$x more problems, each with approximately three times as many test cases on average, making it prohibitively expensive to perform a comprehensive analysis with multiple LLMs.

We start with three single-round prompting methods. First, we use vanilla prompting for performance improvement, instructing the LLM to optimize the correct code while ensuring functional equivalence (Appendix Figure \ref{fig:oneround}(a)). Next, we evaluate a 3-Shot In-context learning baseline, which includes three examples of program refinements along with optimization instructions (Appendix Figure \ref{fig:oneround}(b)). We also assess an improved prompt that includes common Python optimization tricks as predefined strategies, along with the usual optimization instruction (Appendix Figure \ref{fig:oneround}(c)). We then consider two multi-round approaches similar to those in \cite{madaan2024self, nye2021show}, which include a planning or analysis stage, as shown in Figure \ref{fig:tworound} in the Appendix. In the Plan and Refine approach, we prompt the LLM to generate a plan for performance refinement, then prompt it again with this output to implement the proposed refinement. In Analyze and Refine, we prompt the LLM to first analyze the time complexity, then prompt it again to refine the code based on this analysis. 

We also evaluate multi-agent prompting. First, we implement a multi-agent coder-reviewer setup for performance refinement (Figure \ref{fig:multiagent_reviewer}), where a coder refines the base solution and a reviewer provides feedback, followed by another refinement attempt based on this feedback. Additionally, we implement a more elaborate variant with leader-coder-reviewer roles, where the three agents take turns planning, refining, and reviewing code (Figures \ref{fig:multiagent_team_1} and \ref{fig:multiagent_team_2} in the Appendix). Finally, we implement two variants leveraging execution feedback for performance improvements after the LLM attempts to refine the correct code solution. In the first variant (Figure \ref{fig:naive_feedback}), we execute and evaluate the effectiveness of the refinement and verbalize the result (positive if the refinement is faster than the base, negative otherwise), feeding this feedback to the LLM for another refinement attempt. In \tool, as described in Section \ref{sec:meth}, we provide the most time-consuming unit test as feedback to the LLM.

From Table \ref{tab:ablation_prompting}, we observe that the base model generations offer significant performance optimizations over the ground truth. However, the gains with performance-improvement prompting, multi-round, and multi-agent techniques are insignificant and often negative. This can be attributed to cascading LLM reasoning errors, as discussed in Section \ref{sec:intro}. Direct execution feedback produces mixed results, with a \%Opt gain of 5.55\% on HumanEval and a drop of 1.31\% on MBPP. In contrast, \tool results in substantial gains in optimization rate on HumanEval (6.17\% gain in \%Opt) and MBPP (3.06\% gain in \%Opt) problems, validating the higher performance-improvement effectiveness of \tool's execution feedback, in the form of the most expensive unit test.

\begin{table}
    \centering
     \scalebox{0.9}{
    \begin{tabular}{c c c c c c c}
        \toprule
        \multirow{2}*{\textbf{Prompt}} & \multicolumn{3}{c}{\textbf{GPT-3.5}} & \multicolumn{3}{c}{\textbf{Llama 3 70B}} \\
        \cmidrule{2-7}
         &  \textbf{Best@1} & \textbf{Best@8} & \textbf{Best@20} & \textbf{Best@1} & \textbf{Best@8} & \textbf{Best@20}  \\
        \midrule
        \multicolumn{7}{c}{\textbf{HumanEval}} \\
        \midrule
        Base & 54.94 & 78.40 & 84.57 & 75.61 & 84.76 & 89.02 \\
        Testcase Feedback &72.84 & 85.80 & 90.74 & 84.76 & 90.24 & 92.07 \\
        \tool & 68.52 & 90.12 & 93.83 & 85.37 & 93.90 & 95.12 \\
        \midrule
        \multicolumn{7}{c}{\textbf{MBPP (test)}} \\ %\hline
        \midrule
        Base & 66.38 & 76.86 & 79.48 & 65.32 & 77.93 & 81.53 \\
        Testcase Feedback &78.60 & 90.83 & 91.27 & 72.97 & 87.39 & 91.89 \\
        \tool & 73.36 & 88.21 & 92.58 & 68.47 & 88.29 & 93.24\\
        % \midrule
        % \multicolumn{7}{c}{\textbf{APPS (test)}} \\ %\hline
        % \midrule
        % Base & 25.18 & 39.92 & 44.84 & 18.65 & 26.12 & 34.15 \\
        % Testcase Feedback & 29.33 & 44.66 & 49.58 & 20.50 & 30.26 & 39.36 \\
        % \tool & 30.56 & 46.26 & 51.34 & 19.30 & 27.94 & 37.08 \\
        \bottomrule \\
    \end{tabular}}
    
    \caption{\%Correct using GPT-3.5 and Llama 3 70B at $k$ of $1$, $8$ and $20$.}
    \label{tab:correctness}
    \vspace{-5mm}
\end{table}

\subsection{Role of Planning in Correctness Refinement}
\label{subsec:ablation_correct}
We evaluate the effectiveness of \tool's planning step in achieving higher correctness compared to directly using execution feedback. As discussed in Section \ref{subsec:correct_code}, a high correctness rate is essential for generating optimized solutions effectively across a larger proportion of test set problems. \tool is more likely to produce maximally optimal solutions by refining from a larger pool of correct candidate solutions, benefiting from the greater diversity within the seed set. 

We implement the (direct) Testcase Feedback approach as follows: starting with the base LLM generation, we evaluate correctness based on available unit tests and instruct the LLM to refine its solution according to the environment output (Appendix Figure \ref{fig:correctness}(b). This process mirrors how a developer would verify correctness by executing a program to ensure it passes the test suite. In contrast, \tool incorporates an additional planning step based on verbalized environment output. The generated plan is then included in the prompt for refinement in the subsequent step.

Results with these two approaches are shown in Table \ref{tab:correctness} on the HumanEval and MBPP datasets with GPT-3.5 and Llama 3 70B. Both approaches achieve higher correctness than the base LLM. With a $k$ of $1$, we observe that the additional planning step of \tool often leads to slightly lower correctness gain compared to the direct approach. However, with a $k$ of $8$ and $20$, we generally observe higher correctness rate with the planning step. Notably, \tool achieves the highest correctness rates on both benchmarks, underscoring the utility of its additional planning step, suggesting this should also be included in the performance phase given additional token budget.

%, despite the extra computation for it to be effective (higher $K$ and additional tokens to be generated in the planning step).

\section{Related Work}\label{sec:literature}

Many software engineering works have proposed a code-to-code editing formulation for improving code quality in the form of tasks like fixing bugs (\cite{gupta2017deepfix}), performance improving edits (\cite{aman23learning, garg2022deepperf}), improving maintainability (\cite{loriot2022styler, al2022readable}), and security enhancing edits (\cite{{he2023large, perry2023users, tony2023llmseceval, pearce2022asleep, bhatt2023purple}}). Contrary to this approach, we formulate a text-to-code task for our work on runtime efficiency aspect of quality improvements. As programmers continue to rely on prompting LLMs for generating programs for repetitive tasks in software engineering (\cite{dohmke2023sea, feng2024prompting, white2023chatgpt, denny2023conversing}), we opine that it is critical for research on code quality to focus on the prompting stage by studying natural language inputs that describe developer intent or program specifications.

To improve the general LLM output quality (\cite{chiang2024chatbot}) post the pre-training and supervised instruction fine-tuning (\cite{weifinetuned}) stages, recently proposed algorithms like Reinforcement Learning from Human Feedback (\cite{ouyang2022training, stiennon2020learning}) and Direct Preference Optimization (\cite{rafailov2024direct}) that use human preference data have become industry standard (\cite{touvron2023llama}). While one could continue to scale these approaches for improving LLM generated code quality, this would require gathering large-scale preference data for code, which is arguably more difficult and expensive than collecting natural language response preferences. Besides needing extensive number of samples, RL techniques also involve expensive model fine-tuning and are known to be notoriously prone to training instabilities (\cite{casper2023open, wang2024rlhf}). 

Our work builds upon the effectiveness of prior work like \cite{madaan2024self}'s Self-Refine, \cite{nye2021show}'s Scratchpads for LLMs and \cite{chen2023teaching}'s Self-Debug who propose LLM based self-refinements to improve output quality by adding intermediate planning or analysis stages. Our framework is also closely related to \cite{shinn2024reflexion}'s Reflexion who use environment or tool feedback to improve LLM output quality, but focus only on functional correctness in the context of code generation. With \tool, we extend these ideas to improve program runtime efficiency, an aspect that has been largely ignored in favor of functional correctness by prior work. % on improving LLM output quality. %, which is a critical aspect of code quality.

% Our work is most closely related to \cite{aman23learning} who tackle the challenge of suggesting performance-improving code edits using LLMs. We believe obtaining optimal code based on the requirements described in natural language and unit tests is a more impactful problem to solve. Note that we assume access to the set of unit tests $U(x)$ for solving $x$, which is unlike many prior work like Reflexion TODO\cite{} that focus on correctness assuming partial or no access to $U(x)$. Our work focuses on performance of correct code, and we assume access to $U$ along with $x$, which is close to the setup how software is commonly developed in the real world (SDLC?) TODO \cite{}. Unit test generation is another key area where LLMs have been shown to be promising TODO \cite{}. Such advanced automatic and reliable unit test generation becoming widely adopted in real world software development makes the setup of our approach very relevant, making it particularly useful for software optimizations.
\section{Conclusion}\label{sec:conclusion}

We introduced \tool, a novel framework that leverages code execution feedback during the iterative self-refinement stages of LLMs to enhance the runtime-efficiency of generated code. We show that using our approach, open LLMs like Phi-3-mini can achieve code quality comparable to closed LLMs like GPT-4. Our evaluation of \tool on three widely used Python programming benchmarks using both open and closed language models of varying sizes, demonstrates consistent and significant gains in correctness and runtime efficiency. On a sizeable fraction of the test set problems from HumanEval and MBPP, we achieve programs with state-of-the-art runtime efficiency, far exceeding the ground truth reference solutions in several cases with \tool using GPT-4. These findings underscore the importance of integrating execution feedback into the code generation process, highlighting a path forward for more robust and reliable AI-driven software development.

% \newpage
% TODO list from first 3 sections:

% <TODO: Mention the compute needed>
% <TODO: resolve terminology conflict - efficiency, performance, time execution, speed, fast, optimal etc. | similarly there's conflict with programming problems/tasks/description etc. | refine/optimize/perf improve etc.>
% <TODO: Speedup comments>
% <TODO: APPS big picture takeaway>
% <TODO: qualitative example - pull one from appendix>
% <TODO: highlight this aspect: more correct solutions, better it is as you start refining more divers solutions, you'll likely have diverse refinements, and the best(A) > best(A') if A is superset of A'>
% <ablation's APPS results>
% <all links to appendix>
% <sat we ideally want to do planning in perf improving phase too, but token limited>
% <conclusion can inspire future work to explore RAG style perf improvements>

% \newpage
\bibliographystyle{plainnat}
\bibliography{ref}

% \newpage
% \input{sections/neurips_checklist}

\newpage
\appendix
\section{Limitations}\label{sec:limitations}

The challenge of writing efficient and high-quality software with LLMs spans various levels of granularity, from line-level optimizations to multi-class project repositories \cite{shrivastava2023repository}. In our current scope, we focus on generating performant modules or Python functions, which are typically small components of real-world systems. However, addressing this problem comprehensively should ideally involve ensuring architectural design patterns such as minimal redundancy or wasteful computation across the entire scope of a project. 

Another limitation is our focus solely on measuring the runtime performance of LLM-generated code, disregarding memory consumption, which can be crucial in many applications. Future extensions of \tool could prioritize optimizing for both factors or allow users to specify preferences for optimization. Additionally, beyond performance, developers desire attributes like readability, ease of maintenance, security, and harmlessness \cite{bhatt2023purple, singhal2024nofuneval}, which are not within the scope of our current work. While \tool could be adapted to incorporate feedback from different environments or tools evaluating these attributes, achieving a balance in optimizing code generation across these dimensions is non-trivial.

As emphasized by prior research, reliably measuring the runtime performance of code poses challenges \cite{aman23learning}. A piece of code may exhibit varying execution times across different compute environments, even with identical underlying hardware. Unfortunately, tools like the gem5 simulator \cite{binkert2011gem5} do not support the execution of Python programs at the time of our study. To mitigate this, we ensure identical compute environments for each candidate code snippet and run only a single Python program at any given time to minimize effects from concurrent execution. However, averaging execution time measurements from 10 independent runs of each program significantly adds to our execution costs. Future work could explore more efficient methods for reliably measuring runtime, such as determining the instruction count of LLM-generated programs deterministically.
% \newpage
\section{Appendix}
\subsection{Sanitized Benchmarks}\label{a:benchmark}

\begin{table}[h]
    \centering
    \begin{tabular}{ccccccc}
    \toprule
       \multirow{2}*{\textbf{Benchmark}}  & \multicolumn{3}{c}{\textbf{Original}} & \multicolumn{3}{c}{\textbf{Sanitized}} \\
       \cmidrule{2-7}
       & \#Problem &  \#Groundtruth & \#Testcase & \#Problem & \#Groundtruth & \#Testcase\\
       \midrule
        HumanEval & 164 & 1.0 & 9.6 & 164 & 1.0 & 9.6 \\
        MBPP-test & 257 & 1.0 & 3.0 & 257 & 1.0 & 3.0 \\
        APPS-test & 5,000 & 30.4 & 21.2 & 3,249 & 26.9 & 27.4 \\
    \bottomrule \\
    \end{tabular}

    \caption{Details of sanitized benchmarks used in the evaluation. ``\#Problem'' indicates the number of problems in the benchmark, ``\#Groundtruth'' indicates the average number of ground truth programs for each problem, and ``\#Testcase'' indicates the average number of test cases for each problem. We use the sanitized benchmarks for correctness evaluation and the common subsets for time efficiency evaluation.}
    
    \label{tab:benchmark}
\end{table}

We utilize three benchmarks for our evaluation: the HumanEval benchmark developed by \cite{chen21evaluating}, the sanitized version of the MBPP benchmark created by \cite{austin2021program}, and the APPS benchmark established by \cite{hendrycksapps2021}. Detailed information on these benchmarks is presented in Table~\ref{tab:benchmark}, which we directly obtain from the HuggingFace Hub. In our evaluation process, we initially verify whether the provided ground truth programs within each benchmark can successfully pass the associated test cases. Notably, we identify 1,511 instances in the APPS benchmark where the ground truth solutions fail to meet the test case criteria. This discrepancy arises because our evaluation requires exact matches between program outputs and expected outputs in test cases, whereas the APPS benchmark allows for similar matches and may contain incorrect ground truth programs. This observation aligns with previous findings reported by \cite{dou2024stepcoder}. To ensure a fair comparison among the different benchmarks, we exclude the aforementioned 1,751 instances from our analysis, retaining the remaining 3,249 instances for correctness evaluation. Additionally, for assessing time efficiency, we construct subsets from the three benchmarks, comprising problems for which all baseline methods can generate at least one functionally correct program. Time-related data are computed solely on these subsets, rather than the entire benchmark, to maintain consistency in evaluation conditions.

\subsection{Compute}\label{a:compute}

\textbf{LLM inference}: We use the vLLM \cite{kwon2023efficient} library on a node with $16$ Nvidia A100 GPUs for approximately three weeks to complete all the experiments in this work.

\textbf{OpenAI API}: We use the gpt-4-0613 and gpt-3.5-turbo-0125 model endpoints from OpenAI. In total, we required nearly $411$k GPT-4 and $1.5$M GPT-3 requests for all our experiments, contributing to the major costs of this study.

\textbf{Code Execution}: We use $40$ instances of virtual machines (n1-highmem-16 GCP instances), each with $16$ CPUs and $104$ GB RAM for executing all the LLM generated programs generated in our experiments. We employ these instances for roughly four weeks to complete the execution of all the LLM generated programs using different frameworks in our study. Interestingly, unlike most LLM research, gathering this environment feedback tends to be the much costlier bottleneck in our experiments compared to the LLM inference costs. We implement the safeguards prescribed in Section 2.3 of \cite{chen21evaluating} to mitigate the security risks in executing untrusted programs in our environment.

\subsection{Statistical Significance}
\label{a:stat_sig}

Our main results are based on findings in Table \ref{tab:main_result_he_mbpp} and Table \ref{tab:main_result_apps}, where we report that using \tool leads to significant gains in the \%Opt metric compared to the base LLM. For results with the GPT-4 model, we compute the Z-scores to compare \tool's output with that of the base model: $-0.878 (k = 1)$ and $-1.34 (k = 8)$ on HumanEval, $-2.588 (k = 1)$ and $-2.947 (k = 8)$ on MBPP and $-1.8 (k = 1)$ and $-2.675 (k = 8)$ on APPS. The improvement obtained with \tool is thus statistically significant with $\alpha < 0.05$ on the MBPP and APPS problems, and with a lower confidence ($\alpha < 0.2$) on HumanEval which has a smaller number of problems ($164$). %, when $k = 1$.

\subsection{Broader Impact}\label{a:broader_impact}

By enabling LLMs to generate code that is not only functionally correct but also efficient, our work with \tool can significantly accelerate the software development process. This can lead to faster creation of new applications, reduced development costs, and increased innovation across various industries. Frameworks like \tool can potentially empower individuals with less coding experience to leverage LLMs for basic programming tasks. This could lead to a wider pool of software developers and a more inclusive tech landscape. More efficient code translates to lower energy consumption during program execution. This can contribute to a more sustainable software development ecosystem and reduce the environmental impact of the tech industry. Our work demonstrates the ability of \tool to enhance the performance of open LLMs, making them more competitive with closed models. This can foster a more open and accessible environment for LLM development and research. Our work could also offer a promising route to discover novel algorithms to solve long standing problems more efficiently (\cite{romera2024mathematical}). 

While \tool aims to improve code efficiency, it could be misused to automate the generation of efficient harmful or malicious code. For instance, cybercriminals could use optimized code to create more efficient malware or exploit software vulnerabilities more effectively. Incorporating content filtering and malicious code detection algorithms to identify and block harmful code generation can help reduce the risk of such misuse. Optimized code may sometimes introduce new types of vulnerabilities that are difficult to detect. If \tool generates code that is highly efficient but less readable or maintainable, it could lead to challenges in debugging and maintaining software, potentially resulting in unexpected failures or security issues. Addressing this risk requires implementing comprehensive testing and validation, including code reviews, to ensure the generated code maintains high reliability and security standards. More efficient code generation could lead to further automation in software industry, potentially displacing some human programmers. This necessitates discussions on re-skilling initiatives and the evolving nature of jobs in the tech industry.
\newpage
\subsection{Best@20 Results of \tool for All Models}

\begin{table}
    \centering
    \begin{tabular}{c c c c }
        \toprule
        Model &  \textbf{Speedup} & \textbf{Opt\%} & \textbf{Correct\%} \\
        \midrule
        \multicolumn{4}{c}{\textbf{HumanEval}} \\
        \midrule
        Phi3 & 1.81 & 47.56 & 92.68  \\
        Mixtral-8x7B & 1.85 & 42.14 & 87.42 \\
        Command R & 2.01 & 46.63 & 87.12   \\
        Llama3 8B & 2.27 & 43.9 & 87.2    \\ 
        Llama3 70B & 1.84 & 45.73 & 94.51\\
        GPT-3.5 & 2.33 & 43.21 & 93.83 \\
        GPT-4 & 1.98 & 55.21 & 95.71  \\
        \midrule
        \multicolumn{4}{c}{\textbf{MBPP (test)}} \\ %\hline
        \midrule
        Phi3 & 3.40 & 59.91 & 89.22   \\
        Mixtral-8x7B & 3.52 & 47.3 & 87.39  \\
        Command R & 3.21 & 46.26 & 85.9   \\
        Llama3 8B & 3.30 & 43.72 & 86.15  \\ 
        Llama3 70B & 2.88 & 44.59 & 93.24 \\ 
        GPT-3.5 & 4.25 & 75.98 & 92.58  \\
        GPT-4 & 4.16 & 66.96 & 95.65  \\ %\hline
        \midrule
        \multicolumn{4}{c}{\textbf{APPS (test)}} \\ %\hline
        \midrule
        Phi3 & 2.83 & 0.65 & 21.46  \\
        Mixtral-8x7B & 1.75 & 0.59 & 21.42   \\
        Command R & 1.05 & 1.26 & 34.53   \\
        Llama3 8B & 2.86 & 0.95 & 22.44   \\ 
        Llama3 70B & 1.06 & 1.39 & 37.08   \\ 
        GPT-3.5 & 1.24 & 2.06 & 51.34   \\
        GPT-4 & 2.57 & 2.95 & 72.18   \\
        \bottomrule \\
    \end{tabular}
    \caption{\tool results on HumanEval, MBPP and APPS with a $k$ of $20$.}
    \label{tab:main_result_best_20}
\end{table}
% \newpage
% \input{appendix/ablation}
\newpage
\subsection{Examples of Optimal Solutions Generated by LLMs}\label{a:examples}

\begin{lstlisting}[language=python,caption=An optimal solution generated by \tool in HumanEval.,label=lst:humaneval]
'''
Problem:
The FibFib number sequence is a sequence similar to the Fibbonacci sequnece that's defined as follows:
fibfib(0) == 0
fibfib(1) == 0
fibfib(2) == 1
fibfib(n) == fibfib(n-1) + fibfib(n-2) + fibfib(n-3).
Please write a function to efficiently compute the n-th element of the fibfib number sequence.
>>> fibfib(1)
0
>>> fibfib(5)
4
>>> fibfib(8)
24
'''
# Optimal solution generated by PerfCodeGen based on GPT-3.5
def fibfib(n: int):
    if n == 0 or n == 1:
        return 0
    if n == 2:
        return 1
    (a, b, c) = (0, 0, 1)
    for _ in range(3, n + 1):
        (a, b, c) = (b, c, a + b + c)
    return c
# Original ground truth solution in HumanEval:
def fibfib(n):
    if n == 0:
        return 0
    if n == 1:
        return 0
    if n == 2:
        return 1
    return fibfib(n - 1) + fibfib(n - 2) + fibfib(n - 3)
\end{lstlisting}

% \begin{lstlisting}[language=python,caption=An optimal solution generated by \tool in MBPP.,label=lst:mbpp]
% '''
% Problem: Write a python function to find the average of cubes of first n natural numbers.
% '''
% # Optimal solution generated by PerfCodeGen based on GPT-4:
% def solution(n):
%     sum_of_cubes = (n * (n + 1) / 2.0) ** 2
%     return sum_of_cubes / n
    
% # Original ground truth solution in MBPP:
% def solution(n):
%     sum = 0
%     for i in range(1, n + 1):
%         sum += i * i * i
%     return round(sum / n, 6)
% \end{lstlisting}

\newpage
\subsection{Prompts}\label{a:prompts}
We design multiple prompts to evaluate the base performance of LLM and generate runtime-efficient code solutions.

\begin{figure}[h]
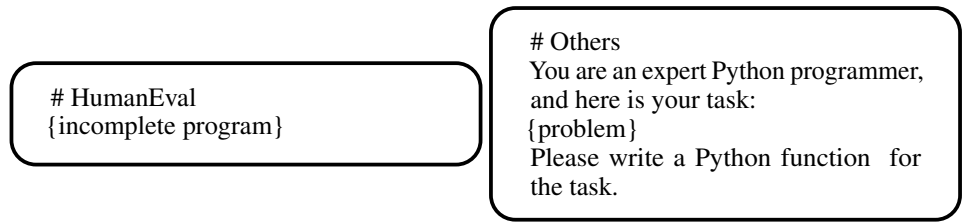

    \begin{minipage}{0.45\textwidth}
    \prompt{\# HumanEval \\
    \{incomplete program\}
    }
    \end{minipage}
    \begin{minipage}{0.45\textwidth}
    \prompt{\# Others \\
    You are an expert Python programmer, and here is your task: \\
    \{problem\}\\
    Please write a Python function {} for the task.
    }
    \end{minipage}
    
    \caption{Base prompt for HumanEval and other benchmarks.}
    \label{fig:base}
\end{figure}

\begin{figure}[t]
    \subfigure[Reflection and Test Case Feedback]{
    \prompt{\textbf{Round 1:} 
    
    Your generated solution for the problem is not correct and cannot pass the following test case: \\
    
    \textit{\{testcase\}} \\
    
    The error message is as follows: \\
    
    \textit{``\{error\}''} \\
    
    Could you please analyze the reason of failure and propose a strategy to modify your solution so that it can pass the above test case?

    \tcbline
    
    \textbf{Round 2 :}
    
    Could you please modify your solution so that it can fulfill the requirements in the problem and pass the test case?
    
    Give your solution as follows. Wrap it with ```python```.
    }
    }
    \subfigure[Test Case Feedback]{
    \prompt{
    Your generated solution for the problem is not correct and cannot pass the following test case: \\
    
    \textit{\{testcase\}} \\
    
    The error message is as follows: \\
    
    \textit{``\{error\}''} \\
    
    Could you please modify your solution so that it can fulfill the requirements in the problem and do not have any syntax error?
    
    Give your solution as follows. Wrap it with ```python```.
    }
    }
    
    \caption{The two correctness prompts discussed in the paper. \tool uses the reflection and test case feedback prompt in (a).}
    \label{fig:correctness}
\end{figure}

\begin{figure}[t]
    \centering
    \subfigure[Perf Improvement Prompt]{
    \prompt{
    Good job! You generated the correct solution for the problem! Now let's step further and optimize the time performance of the solution.
    
    Based on the correctly generated solution, could you please refine it so that it consumes less time in the execution?

    Please make sure your refined solution is functionally equivalent with the original solution and do not change the input-output format and the name of the major components.

    Give your solution as follows. Wrap it with ```python```.
    }
    }
    \subfigure[In-context learning (few-shot)]{
    \prompt{
    Good job! You generated the correct solution for the problem! Now let's step further and optimize the time performance of the solution.
    
    Here are some examples of optimization: \\
    
    \textit{\{demo\}} \\
    
    Based on the correctly generated solution and the above examples, could you please refine it so that it consumes less time in the execution?

    Please make sure your refined solution is functionally equivalent with the original solution and do not change the input-output format and the name of the major components.

    Give your solution as follows. Wrap it with ```python```.
    }
    }
    \subfigure[Pre-defined Strategies]{
    \prompt{
    Good job! You generated the correct solution for the problem! Now let's step further and optimize the time performance of the solution.
    
    Here are some commonly used strategies for optimization:
    
    1. Do not generate useless testing code.
    
    2. Use builtin functions and libraries instead of importing third-party libraries.
    
    3. Use list and dict comprehension to avoid loops.
    
    4. Eliminate unnecessary variable definitions, function definitions and print statements.
    
    5. Optimize the loops to avoid unnecessary iterations.
    
    6. Use local variables instead of global variables.
    
    7. Use multiple assignments in one statement instead of in multiple statements.
    
    8. Make use of generators.
    
    Based on the correctly generated solution and the above strategies, could you please refine it so that it consumes less time in the execution?

    Please make sure your refined solution is functionally equivalent with the original solution and do not change the input-output format and the name of the major components.

    Give your solution as follows. Wrap it with ```python```.
    }
    }
    \caption{The single-round runtime performance improving prompts.}
    \label{fig:oneround}
\end{figure}

\begin{figure}[t]
    \centering
    \subfigure[Plan and Refine]{
    \prompt{
    \textbf{Round 1:}
    
    Good job! You generated the correct solution for the problem! Now let's step further and optimize the time performance of the solution.
    
    Could you please propose a strategy to optimize the generated solution so that it consumes less time in the execution?

    \tcbline
    \textbf{Round2:}

    Please implement the strategy you proposed for the previous solution and generate the optimized solution.

    Please make sure your refined solution is functionally equivalent with the original solution and do not change the input-output format and the name of the major components.

    Give your solution as follows. Wrap it with ```python```.
    }
    }
    \subfigure[Analyze and Refine]{
    \prompt{
    \textbf{Round 1:}
    
    Good job! You generated the correct solution for the problem! Now let's step further and optimize the time performance of the solution.
    
    Could you please analyze the time complexity of the solution?

    \tcbline

    \textbf{Round 2:}

    Now given the time complexity you estimated, could you please optimize your solution by reducing its time complexity?

    Please make sure your refined solution is functionally equivalent with the original solution and do not change the input-output format and the name of the major components.

    Give your solution as follows. Wrap it with ```python```.
    }
    }
    \caption{The multi-round runtime performance improving prompts.}
    \label{fig:tworound}
\end{figure}

\begin{figure}[t]
    \centering
    \prompt{
    \textbf{Round 1: }
    Good job! You generated the correct solution for the problem! Now let's step further and optimize the time performance of the solution.
    
    Based on the correctly generated solution, could you please refine it so that it consumes less time in the execution?

    Please make sure your refined solution is functionally equivalent with the original solution and do not change the input-output format and the name of the major components.

    Give your solution as follows. Wrap it with ```python```.

    \tcbline
    \textbf{Round 2 (Negative Feedback): }

    We tested your optimized solution and found that your optimized solution runs slower than the previous version. The following is the execution time for both solutions:
    
    Original solution:\\
    
    \textit{\{ori\_time\}}\\
    
    Your optimized solution:\\
    
    \textit{\{opt\_time\}}\\
    
    Based on the execution time and the original solution, could you please propose another optimized solution so that it consumes less time than the original solution?

    Please make sure your refined solution is functionally equivalent with the original solution and do not change the input-output format and the name of the major components.

    Give your solution as follows. Wrap it with ```python```.

    \tcbline
    \textbf{Round 2 (Positive Feedback): }

    We test your optimized program. Great! It runs faster than the original program.
    
    The following is the execution time for both solutions:
    
    Original solution:\\
    
    \textit{\{ori\_time\}}\\
    
    Your optimized solution:\\
    
    \textit{\{opt\_time\}}\\
    
    Can you refine the current optimized version as follows and provide a more efficient one?

    Please make sure your refined solution is functionally equivalent with the original solution and do not change the input-output format and the name of the major components.

    Give your solution as follows. Wrap it with ```python```.
    }
    \caption{Direct Execution Feedback prompt.}
    \label{fig:naive_feedback}
\end{figure}

\begin{figure}
    \centering
    \prompt{
    \textbf{Round 1: }
    Good job! You generated the correct solution for the problem! Now let's step further and optimize the time performance of the solution.
    
    Based on the correctly generated solution, could you please refine it so that it consumes less time in the execution?

    Please make sure your refined solution is functionally equivalent with the original solution and do not change the input-output format and the name of the major components.

    Give your solution as follows. Wrap it with ```python```.

    \tcbline

    \textbf{Round 2:}

    We tested your optimized program and found that the following test case costs the most time in execution.\\
    
    \textit{\{testcase\}}\\
    
    Could you please refine your optimized program according to the test case below?

    Please make sure your refined solution is functionally equivalent with the original solution and do not change the input-output format and the name of the major components.

    Give your solution as follows. Wrap it with ```python```.
    }
    \caption{\tool Testcase Feedback prompt used.}
    \label{fig:testcase_feedback}
\end{figure}

\begin{figure}[t]
    \centering
    \prompt{
    \textbf{Round 1 (Coder):}

    Good job! You generated the correct solution for the problem! Now let's step further and optimize the time performance of the solution.
    
    Based on the correctly generated solution, could you please refine it so that it consumes less time in the execution?

    Please make sure your refined solution is functionally equivalent with the original solution and do not change the input-output format and the name of the major components.

    Give your solution as follows. Wrap it with ```python```.

    \tcbline

    \textbf{Round 2 (Reviewer):}
    You are in a discussion group, aiming to optimize a Python program so that it runs faster.
    
    The original Python program is:\\
    
    \textit{\{program\}}\\
    
    An optimized Python program given by your group member:\\
    
    \textit{\{opt\_program\}}\\
    
    You are a reviewer. Based on your knowledge, can you check whether the optimized program given by your group member runs faster than the original program?
    
    You response should follow the following rules:
    
    1. Begin your response with [Agree] if you think the optimized program really runs faster than the original one and [Disagree] if not or there exists some syntax errors in the optimized program.
    
    2. Provide your comments after the keyword ``Comment:'' to explain your decision.
    
    3. Give suggestions for further improvements in comments.
    
    Give your response here:

    \tcbline

    \textbf{Round 3 (Coder):}

    A code reviewer carefully reviewed your optimized code solution. He thinks that your optimization is \textit{\{decision\}}.
    
    Please refer to his comments as follows to further refine your optimized program.
    
    Comments:\\
    
    \textit{\{comment\}}\\

    Please make sure your refined solution is functionally equivalent with the original solution and do not change the input-output format and the name of the major components.

    Give your solution as follows. Wrap it with ```python```.
    }
    \caption{The Multi-Agent (Coder - Reviewer) prompt.}
    \label{fig:multiagent_reviewer}
\end{figure}

\begin{figure}[t]
    \centering
    \prompt{
    \textbf{Round 1 (Leader):}

    You are a team leader and your team is working on the code of the following problem:\\
    
    \textit{\{problem\}}\\
    
    Here is the correct solution your team wrote before:\\
    
    \textit{\{program\}}\\
    
    Given the Python program, can you make a plan about how to optimize it step by step?
    
    Give your plan here:

    \tcbline
    \textbf{Round 2 (Coder):}

    Given the previous Python program, your leader has proposed the following plan to optimize it:\\
    
    \textit{\{plan\}}\\
    
    Following the above plan, can you write an optimized version of the original Python program?

    Please make sure your refined solution is functionally equivalent with the original solution and do not change the input-output format and the name of the major components.

    Give your solution as follows. Wrap it with ```python```.

    \tcbline
    \textbf{Round 3 (Reviewer):}

    You are a code reviewer in a team and your team is trying to optimize the following Python program so that it runs faster.
    
    Original Python program:\\
    
    \textit{\{program\}}\\
    
    Given the Python program, your leader has proposed the following plan to optimize it:\\
    
    \textit{\{plan\}}\\
    
    The coder in your team implemented an optimized version of program as follows. \\
    
    \textit{\{opt\_program\}} \\
    
    Based on your knowledge, can you check whether the optimized program given by your group member runs faster than the original program?
    
    You response should follow the following rules:
    
    1. Begin your response with [Agree] if you think the optimized program really runs faster than the original one and [Disagree] if not or there exists some syntax errors in the optimized program.
    
    2. Provide your comments after the keyword \"Comment:\" to explain your decision.
    
    3. Give suggestions for further improvements in comments.
    
    Give your response here:
    }
    \caption{The first part of Multi-Agent w/ Team (Leader - Coder - Reviewer) prompt.}
    \label{fig:multiagent_team_1}
\end{figure}

\begin{figure}[t]
    \centering
    \prompt{
    \textbf{Round 4 (Leader):}

    A coder in your team implement an optimized version of program according to your plan: \\
    
    \textit{\{opt\_program\}} \\
    
    A code reviewer reviewed the optimized version and thinks that the optimization is \textit{\{decision\}}.
    
    He provides the following comments: \\
    
    \textit{\{comment\}} \\
    
    Could you please refine your plan so that we can further improve the efficiency of the program?
    
    Give you plan here:

    \tcbline
    \textbf{Round 5 (Coder):} 

    After your optimized program is reviewed by another group member, your leader modifies his plan and gives a new plan here: \\
    
    \textit{\{plan\}} \\
    
    Could you please write a new optimized version of the program based on the new plan?

    Please make sure your refined solution is functionally equivalent with the original solution and do not change the input-output format and the name of the major components.

    Give your solution as follows. Wrap it with ```python```.
    }
    \caption{The second part of Multi-Agent w/ Team (Leader - Coder - Reviewer) prompt.}
    \label{fig:multiagent_team_2}
\end{figure}

\end{document}